\documentclass[%
 reprint,
nofootinbib,
 amsmath,amssymb,
 aps,
 longbibliography,
]{revtex4-2}

\usepackage{graphicx}
\usepackage{dcolumn}
\usepackage{bbm}
\usepackage{physics}
\usepackage{braket}
\usepackage{mathtools}
\usepackage{graphicx} 
\usepackage{hyperref}
\usepackage{cleveref}
\usepackage{etoolbox}
\usepackage{xcolor}
\usepackage{siunitx}
\newcommand{\rmi}{\ensuremath{\mathrm{i}}}
\newcommand{\rme}{\ensuremath{\mathrm{e}}}
\newcommand{\rmd}{\ensuremath{\mathrm{d}}}

\newcommand{\nnl}{\nonumber\\}

\newcommand{\id}{\ensuremath{\mathbbm{1}}}

\begin{document}


\title{State Swapping via Semiclassical Gravity}

\author{Praveer K. Gollapudi}
\email{praveer.gollapudi@uni-jena.de}
\author{M.~Kemal D\"oner}%
\email{kemal.doner@uni-jena.de}
\author{Andr\'e Gro{\ss}ardt}
\email{andre.grossardt@uni-jena.de}
\affiliation{Institute for Theoretical Physics, Friedrich Schiller University Jena, Fr\"obelstieg 1, 07743 Jena, Germany}%

\date{\today}

\begin{abstract}
 The prediction of non-local phenomena is a key attribute of quantum mechanics that distinguishes it from classical theories. It was recently suggested that state swapping is one such effect that a fundamentally classical gravitational field cannot give rise to. The reasoning was that the phenomenology of classical fields is restricted to Local Operations and Classical Communication~(LOCC) whereas swapping is a non-LOCC effect, so gravity must not be a classical channel if it can give rise to such a phenomenon. Here, we show that the Hamiltonian of semiclassical gravity results in an evolution of coherent states that is virtually indistinguishable from swapping them, both theoretically and experimentally. Our results indicate that drawing a definitive conclusion about the nature of gravity based solely on its ability to swap a pair of coherent states would be premature. We discuss the implications of these results regarding proposed experimental tests for the quantum nature of gravity and clarify some conceptual issues.
\end{abstract}

\maketitle


\section{Introduction}

In almost a century of quantum gravity (QG) research, only minimal progress has been made toward finding empirical evidence for or against any of the candidate theories suggested. Progress in quantum technology and high-fidelity laboratory experiments has recently shifted some attention to the idea of finding an experimental answer, at least, to the much broader question of whether gravity should be quantized at all.

Past discussions in this field focused on the alternative hypothesis of semiclassical gravity~\cite{mollerTheoriesRelativistesGravitation1962,rosenfeldQuantizationFields1963}, where quantum matter is coupled to a classical spacetime geometry via the semiclassical Einstein equations, i.e., Einstein's field equations with the expectation value $\bra{\Psi}\hat{T}_{\mu\nu}\ket{\Psi}$ of stress-energy in some quantum state $\ket{\Psi}$ of the matter fields acting as a gravitational source term. We refer to this model as semiclassical Einstein gravity (SCEG) to distinguish it from other, more recent hybrid quantum-classical models. Attempts to exclude SCEG based on arguments of theoretical consistency~\cite{eppleyNecessityQuantizingGravitational1977,pageIndirectEvidenceQuantum1981} are inconclusive~\cite{kibbleSemiClassicalTheoryGravity1981,mattinglyWhyEppleyHannah2006,albersMeasurementAnalysisQuantum2008,kentSimpleRefutationEppley2018}. Nonetheless, the consistent semiclassical coupling of gravity and quantum matter requires fundamental changes for quantum mechanics (QM), specifically regarding the measurement process~\cite{kibbleSemiClassicalTheoryGravity1981,giuliniCouplingQuantumMatter2022}. Several promising avenues toward an experimental test~\cite{carlipQuantumGravityNecessary2008,giuliniGravitationallyInducedInhibitions2011,yangMacroscopicQuantumMechanics2013,grossardtOptomechanicalTestSchrodingerNewton2016,haineSearchingSignaturesQuantum2021a,grossardtSelfgravitationalDephasingQuasiclassical2024} of semiclassical gravity in laboratory settings have been explored. All of them consider the nonrelativistic regime, where SCEG can be shown~\cite{giuliniSchrodingerNewtonEquationNonrelativistic2012,bahramiSchrodingerNewtonEquationIts2014,giuliniCouplingQuantumMatter2022} to result in a nonlinear Schr\"odinger equation, commonly referred to as the Schr\"odinger-Newton (SN) equation.

A more recent path of research, instead, is concerned with the question of whether gravity constitutes a quantum or a classical channel in the sense of quantum information science. If the spatially separated subsystems of a bipartite quantum system, living in a Hilbert space $\mathcal{H} = \mathcal{H}_1 \otimes \mathcal{H}_2$, interact only gravitationally over a limited time interval, this interaction can be modeled by a map $\Lambda: \mathrm{End}(\mathcal{H}) \to \mathrm{End}(\mathcal{H}), \, \hat{\rho}_i \mapsto \hat{\rho}_f$ between initial and final density operators. To have a consistent probabilistic interpretation, this map must be completely positive and trace-preserving (CPTP); a class of maps often referred to as ``channels'' in the quantum information literature. The question of the ``quantumness'' of the gravitational field is then boiled down to the question of whether the map $\Lambda$ (for arbitrary input states) can be decomposed into a product of local operations and classical communication (LOCC)~\cite{plenioIntroductionEntanglementMeasures2007}---in which case gravity would be considered as ``classical'' whereas the attribute ``quantized'' applies to any model that is not classical in this sense. Here, local operations are to be understood in terms of subsystem locality, i.e. unitary operators $U_1 \otimes \id$ or $\id \otimes U_2$ that act trivially on one of the subsystems. Classical communication includes such operations that are functions of local measurement outcomes.

The idea of testing whether gravity can be modeled as a LOCC channel has been introduced by Kafri et al.~\cite{kafriClassicalChannelModel2014b} who propose a classical channel model for Newtonian gravity. Their model is refuted by existing empirical data~\cite{altamiranoGravityNotPairwise2018} but has inspired other models~\cite{tilloySourcingSemiclassicalGravity2016}. It has also resulted in the proposal of experimental tests of gravity-induced entanglement as alleged evidence for the quantization of the gravitational field~\cite{boseSpinEntanglementWitness2017b,marlettoGravitationallyInducedEntanglement2017}.

The foundation of these suggestions is the inability of a LOCC channel to map separable initial states to entangled states~\cite{plenioIntroductionEntanglementMeasures2007}. The set of non-LOCC maps is, however, strictly larger than the set of entangling maps, an important example being the swap map
\begin{eqnarray}\label{eqn:swap}
    \hat{\rho}_i = \hat{\rho}_1 \otimes \hat{\rho}_2 \mapsto \hat{\rho}_2 \otimes \hat{\rho}_1 = \hat{\rho}_f \,.
\end{eqnarray}
Although this map is non-entangling, taking a separable initial state to a separable final state, it cannot be reduced to local operations and classical communication. In a generalization of this thought, where $\hat{\rho}_{1,2}$ are both coherent states of harmonic oscillators, Lami et al.~\cite{lamiTestingQuantumnessGravity2024} recently proposed an experimental test for non-LOCCness that works without generating entanglement.

A matter of active debate is whether the distinction between LOCC and non-LOCC is a meaningful characterization of classical versus quantized gravity. This characterization has been challenged by hybrid models~\cite{hallTwoRecentProposals2018,donerGravitationalEntanglementEvidence2022} which are capable of entangling two particles despite having features that, arguably, are better described as classical than quantum.

Here, we intend to challenge this characterization even further by establishing that SCEG as the historically most considered hybrid approach is indeed \emph{not} a LOCC theory. We achieve this by showing that a swap
\begin{equation} \label{eq:swap}
    \ket{\alpha}_1 \otimes \ket{\beta}_2 \mapsto \ket{\beta}_1 \otimes \ket{\alpha}_2
\end{equation}
of coherent harmonic oscillator states can be implemented with a semiclassical gravitational interaction in the same way it can be implemented in QG. We further discuss how the differences between QG and SCEG shape up once superpositions of coherent states---and therefore entanglement---enter the picture.


Our paper is structured as follows: in section \ref{sec:QG} we use the  Heisenberg equations of motion(EoM) to show that an approximation of the effective QG Hamiltonian leads to the swapping of a pair of coherent states. We then apply certain considerations to the \emph{full} effective QG Hamiltonian to obtain identical equations---and thus identical physics---to those seen in the approximate case.  In section \ref{sec:SCG}, we present the EoM for SCEG and show that applying the same considerations as before leads yet again to equations that predict a swap of coherent states. We conclude the paper with a discussion of our results in the context of the discourse over quantum gravity.

\section{State Swapping in Quantum Gravity}\label{sec:QG}

Consider a pair of non-interacting oscillators, initially in coherent states $\ket{\alpha}$ and $\ket{\beta}$, with identical frequencies and masses. Their combined wave function is given by
\begin{equation*}
    \ket{\psi_0} = \ket{\alpha}_1 \otimes \ket{\beta}_2
\end{equation*}
whereas the total Hamiltonian of the system is given by
\begin{equation*}
    H = H_1 + H_2 = \sum_{i=1,2} \left( \frac{p_i^2}{2 m} + \frac{1}{2} m\omega^2 x_i^2 \right)
\end{equation*}
where $1$ and $2$ label the respective particles. Regardless of the concrete approach to quantum gravity, the low-energy behaviour is generally assumed to be correctly modeled by a perturbative quantization of the metric field, which in the non-relativistic limit results in an interaction Hamiltonian that is simply a Newtonian pair potential~\cite{christodoulouLocallyMediatedEntanglement2023}. That is, assuming that the oscillators under consideration interact only via the quantized gravitational force and the origins of their coordinates $x_1$ and $x_2$ are a distance $d$ apart, the total Hamiltonian is
\begin{align*}
    H^q &= H_1 + H_2 + H_I 
    = H_1 + H_2 - \frac{G m^2}{|d- (x_1-x_2)|} \,.
\end{align*}
If $d \gg |x_1-x_2|$, expanding the denominator of the last term up to second order and discarding~\cite{carneyTabletopExperimentsQuantum2019} those resulting terms which are bi-linear and bi-quadratic in $x_1$ and $x_2$ yields
\begin{subequations} \label{eq:pre_rwa}
\begin{align}
        H^q&=  \sum_{i=1,2}\left[  \frac{p_i^2}{2 m} + \frac{1}{2} m\omega^2 x_i^2 \right]+2 \lambda x_1 x_2 
        \label{eq:pre_rwa_a}\\
        &=\sum_{i=1,2} [\hbar \omega a_i^\dag a_i] + \hbar\omega_g(a_1 a_2^\dag + a_1^\dag a_2 + a_1 a_2 + a_1^\dag a_2^\dag) \label{eq:pre_rwa_b}
\end{align}
\end{subequations}
where $\lambda = G m^2/d^3$ and $\omega_g = \lambda/(m \omega)$.

\subsection{Rotating wave approximation}
The rotating wave approximation (RWA)~\cite{wangBridgingGapJaynes2015,burgarthTamingRotatingWave2024,fujiiIntroductionRotatingWave2017} applies to such Hamiltonians as $H^q$ when the condition $\omega_g \ll \omega$ is met, whereupon the final two terms in \eqref{eq:pre_rwa_b} may be dropped. Justifications for doing so are varied, but the conventional one involves the interaction picture. With this approximation, the Hamiltonian $H^q$ may be simplified to
\begin{subequations}
\begin{align}
H^q_{\text{RWA}}& = \sum_{i=1,2} [\hbar \omega a_i^\dag a_i ]+ \hbar \omega_g (a_1 a_2^\dag + a_1^\dag a_2)\\
&= \sum_{i=1,2} \left[\frac{p_i^2}{2 m} + \frac{1}{2} m\omega^2 x_i^2 \right]+ \lambda \left[ x_1 x_2 + \frac{p_1 p_2}{(m \omega)^2} \right]
\end{align}
\end{subequations}
It must be emphasized that when the oscillator frequencies are identical, as they are here, the credibility of the RWA hinges on  the relative strength of the coupling~\cite{wangBridgingGapJaynes2015,zuecoQubitoscillatorDynamicsDispersive2009}.
Introducing the normal mode coordinates
\begin{equation}\label{eqn:normal-mode-coords}
    x_+ = \frac{x_1 + x_2}{\sqrt{2}}, \quad x_- = \frac{x_1 - x_2}{\sqrt{2}}
\end{equation}
results in a separable Hamiltonian which is easier to solve exactly. Furthermore, a product of coherent states in the original coordinates is mapped to a similar state in the normal mode coordinates and vice-versa: 
\begin{subequations}\label{eq:cs_tran}
\begin{align}
    \ket{\alpha}_1 \otimes \ket{\beta}_2 & = \Ket{\frac{\alpha + \beta}{\sqrt{2}}}_+ \otimes \Ket{\frac{\alpha-\beta}{\sqrt{2}}}_- \label{eq:cs_tran-a}\\
    \ket{a}_+ \otimes \ket{b}_- & = \Ket{\frac{a + b}{\sqrt{2}}}_1\otimes \Ket{\frac{a-b}{\sqrt{2}}}_2 \,.\label{eq:cs_tran-b}
\end{align}
\end{subequations}
In the normal mode coordinates, the Hamiltonian is
\begin{equation}
H^q_{\text{RWA}} = \sum_{j=+,-}k_j\left(\frac{p_j^2}{2m} + \frac{1}{2} m \omega^2 x_j^2\right) \label{eq:rwa_pm} 
\end{equation}
where $k_\pm = 1 \pm \dfrac{\omega_g}{\omega} \equiv 1 \pm \delta$.
This is a Hamiltonian of a pair of decoupled oscillators, whose action on a product of coherent states is easy to calculate in the Schrödinger picture~\cite{gerryIntroductoryQuantumOptics2023}, an approach presumably taken by previous authors~\cite{lamiTestingQuantumnessGravity2024} but one which is not applicable to the semi-classical case.

Instead, we solve the the equations of motion in the Heisenberg picture, which provides an exact solution for the Hamiltonian above---serving as a consistency check---as well as for the semi-classical one, given appropriate approximations (note that due to its non-linearity the full SN equation usually cannot be solved analytically). The key insight underpinning our calculations is that the Hamiltonians under consideration preserve the coherence of a system of oscillators in initially coherent states, charecterized by the condition
\begin{equation}\label{eq:coherence}
   [\hat{a_i},[\hat{a_j},H_{ij}]]= \frac{\partial }{\partial \hat{a_i}^\dag}[\hat{a_j},H_{ij}] = 0 \quad \forall \ i 
\end{equation}
where $\hat{a}_i$ and $\hat{a}^\dag_i$ are the annihilation and creation operators of oscillator $i$. 
In case this condition is met, the time evolution of each oscillator is condensed into the variation of the displacement parameter--- the $\alpha$ in $\ket{\alpha}$--- for that oscillator. The details of the calculation are shown in Appendix \ref{App:WFE}, which require the evolution of the first and second moments calculated from the Hamiltonian in the Heisenberg picture. 

For $H^q_{\text{RWA}}$, solving the equations of motion for the first and second moments (cf. Appendix \ref{App:EoM}) leads to
\begin{subequations}\label{eq:eom_rwa}
\begin{align}
 \braket{x_+(t)} &= \sqrt{\frac{2 \hbar}{m \omega}} (a_R \cos \omega_+ t +  a_I \sin \omega_+ t) \label{eq:eom_rwa-a}\\
\braket{p_+(t)}&=\sqrt{2 \hbar m \omega}(a_I \cos \omega_+ t - a_R \sin \omega_+ t)\\
\Delta x_+^2(t)&= \frac{\hbar}{2m \omega} = \Delta x_-^2(t)\label{eq:eom_rwa-c}\\
 \braket{x_-(t)} &= \sqrt{\frac{2 \hbar}{m \omega}} (b_R \cos \omega_- t +  b_I \sin \omega_- t)\\
\braket{p_-(t)}&=\sqrt{2 \hbar m \omega}(b_I \cos \omega_- t - b_R \sin \omega_- t)
\end{align}
\end{subequations}
where, $\omega_\pm =\omega k_\pm $ and the parameters $a= a_R+\rmi a_I$ and $b = b_R+\rmi b_I$ follow from \eqref{eq:cs_tran-a}.

The width of each state remains constant in time, indicating that coherent states remain coherent, a fact corroborated by $H^q_{\text{RWA}}$ satisfying \eqref{eq:coherence}. Based on the equations above, we find that the initial states evolve as follows:
$$
\ket{a}_+ \otimes \ket{b}_- \mapsto \ket{a \rme^{-\rmi \omega_+ t}}_+ \otimes \ket{b \rme^{-\rmi \omega_- t}}_-
$$
This is the expected evolution of a (pair of decoupled) coherent state(s) under a quantum harmonic potential, which is precisely what $H^q_{\text{RWA}}$ represents. 
Transforming back into the original coordinates, we obtain
\begin{align} \label{eq:rwa_evo}
    \ket{\alpha}_1 \otimes \ket{\beta}_2 \mapsto  &\Ket{\rme^{-\rmi\omega t}(\alpha \cos \omega_g  t -\rmi \beta \sin \omega_g  t)}_1 \otimes \nnl
    &\Ket{\rme^{-\rmi\omega t}(\beta \cos \omega_g  t -\rmi \alpha \sin \omega_g  t)}_2
\end{align}
At time $T$ such that $\omega_g T=\pi/2$,
\begin{equation*}
    \ket{\psi_T}= \ket{- \rmi \rme^{-\rmi \omega T} \beta}_1 \otimes\ket{- \rmi \rme^{-\rmi \omega T} \alpha}_2
\end{equation*}
which effectively swaps the states to within a phase \cite{lamiTestingQuantumnessGravity2024}.

\subsection{Quantum Gravity (full quadratic Hamiltonian)}\label{sec:qg-full}
Consider again the full effective Hamiltonian  $H^q$ \eqref{eq:pre_rwa_a}, which in the $x_\pm$ coordinates~\eqref{eqn:normal-mode-coords}, has the form
\begin{equation}\label{eq:qg}
    H^q 
    = \sum_{j=+,-} \left[\frac{p_j^2}{2 m} + \frac{1}{2} m K_j ^2 \omega^2 x_j^2 \right] 
\end{equation}
where $K_\pm = \sqrt{1 \pm 2 \delta } $. 
Note the structural difference between \eqref{eq:rwa_pm} and \eqref{eq:qg}: the former has a prefactor $k_j$ on both  position and momentum terms of the oscillator Hamiltonian, whereas the latter has a prefactor $K_j^2$ on the position terms only. A direct consequence of this difference is the violation of condition \eqref{eq:coherence}, showing that the QG interaction does not exactly preserve the coherence of initially coherent states. Nevertheless, applying the condition $\omega_g \ll \omega$ to the EoM reproduces the physics contained in \eqref{eq:swap} for all practical purposes.

To that end, consider the equations generated by $H^q$ :
\begin{subequations}\label{eq:qg_eom}
\begin{align} 
\braket{x_+(t)} &= \sqrt{\frac{2 \hbar}{m \omega}} (a_R \cos \Omega_+ t +  \frac{1}{K_+} a_I \sin \Omega_+ t)\label{eq:simi11}    \\
\braket{p_+(t)}&=\sqrt{2 \hbar m \omega}(a_I \cos \Omega_+ t -K_+ a_R \sin \Omega_+ t)\\
\Delta x_+^2(t)& = \frac{\hbar}{2m \omega}\left(\cos^2\Omega_+ t + \frac{1}{K_+^2}  \sin^2 \Omega_+ t \right) \\
\braket{x_-(t)} &= \sqrt{\frac{2 \hbar}{m \omega}} (b_R \cos \Omega_- t +  \frac{1}{K_-} b_I \sin \Omega_- t)   \label{eq:simi21} \\
\braket{p_-(t)}&=\sqrt{2 \hbar m \omega}(b_I \cos \Omega_- t -K_- b_R \sin \Omega_- t)\\
\Delta x_-^2(t)& = \frac{\hbar}{2m \omega}\left(\cos^2\Omega_- t + \frac{1}{K_-^2} \sin^2 \Omega_- t \right) 
\end{align}
\end{subequations}
where $\Omega_+ = \omega K_+$ and $\Omega_- =  \omega K_-$. When $\omega_g\ll\omega$,
$$
K_\pm = \sqrt{1+2\delta} \approx 1 + \delta + \mathcal{O}(\delta^2)
$$

For a typical trapped-ion oscillator with $^{40}\text{Ca}^+$ ions~\cite{alonsoGenerationLargeCoherent2016,kienzlerObservationQuantumInterference2016}, assuming $d\approx \SI{e-10}{\meter}$ (the lattice spacing of $^{40}$Ca) yields $\omega_g \approx \SI{e-12} {\hertz}$, with $\omega \sim \SI{1}{\mega\hertz}$ for the cited works. As such, $\delta$ may be completely disregarded as a correction to the overall amplitude, but not to the local phase since that would lead to loss of information at late times. One must instead write 
$$
 \Omega_\pm t \approx \omega t \left(1  \pm  \delta\right)
$$
because at $t \sim 1/\omega_g $--- our timescale of interest--- we have $ \delta \omega t \sim \pi$, resulting in significant effects to the sine/cosine terms. At much later times, the $\mathcal{O}(\delta^2)$ corrections should also be accounted for; effects in this regime, however, are beyond the scope of this work.

The upshot of this discussion is that we write $K_\pm \approx k_\pm $ within the phases and $K_\pm \approx  1$ outside. Adjusting the EoM accordingly results in a set identical to \eqref{eq:eom_rwa}, which is unsurprising because $\omega_g \ll\omega$ is the key condition for the validity of the RWA. Applying this condition directly to the QG EoM allowed us to reproduce the EoM obtained in the RWA without the need for conservation arguments or the interaction picture.

\section{Semiclassical Gravity}\label{sec:SCG}
We finally consider the case of semiclassical gravity, the hypothesis that the spacetime geometry sourced by quantum matter is shaped by the \emph{average} distribution of energy and momentum:
$$
{G}_{\mu \nu} = \kappa \braket{\hat{T}_{\mu \nu}}
$$
where ${G}_{\mu \nu}$ is the Einstein tensor, $\kappa = 8 \pi G / c^4$, and $\hat{T}$ the stress-energy tensor operator for the quantum matter fields. The non-relativistic version of this equation is the Schrödinger-Newton equation~\cite{giuliniSchrodingerNewtonEquationNonrelativistic2012,bahramiSchrodingerNewtonEquationIts2014}, which in the case of two particles and in situations where self-energy corrections are negligible can be expressed as a Schrödinger equation with a modified Newtonian pair potential. The total Hamiltonian, including a quadratic potential for each particle, reads
\begin{align*}\label{eq:H_sc}
        H^{sc} &= H + V_{12}+V_{21}\\
    &=\sum_{i=1,2}\left[  \frac{p_i^2}{2 m_i} + \frac{1}{2} m\omega_i^2 x_i^2 \right]- \frac{G m^2}{|d- (x_1-\braket{x_2})|} \\
    &- \frac{G m^2}{|d- (\braket{x_1}-x_2)|}
\end{align*}
The pair of oscillators are represented by the same Hamiltonian $H$ as in the QG case, but the interaction between them is modeled by two distinct terms, describing the gravitational attraction of the first particle in the potential of the second and vice versa. If $d \gg |x_i-\braket{x_j}|$, expanding the denominator of $V_{ij}$ will lead to a Hamiltonian similar to, but distinct from, the QG one in~\eqref{eq:pre_rwa_a}:
\begin{subequations}
\begin{align}
    H^{sc}& =  \sum_{i=1,2} \left[\frac{p_i^2}{2 m} + \frac{1}{2} m\omega^2 x_i^2 \right]+ 2\lambda x_1 \braket{x_2}  + 2\lambda x_2 \braket{x_1}\\
    & = \sum_{j=+,-} \left[\frac{p_j^2}{2 m} + \frac{1}{2} m\omega^2 x_j^2 \right]+ 2\lambda x_+ \braket{x_+} - 2\lambda x_- \braket{x_-}
\end{align}
\end{subequations}
The equations of motion resulting from this Hamiltonian are identical to \eqref{eq:qg_eom} except for the widths $\Delta x_\pm$:
\begin{subequations}\label{eq:sc_eom}
\begin{align}
\braket{x_+(t)} &= \sqrt{\frac{2 \hbar}{m \omega}} (a_R \cos \Omega_+ t +  \frac{1}{K_+} a_I \sin \Omega_+ t) \label{eq:simi12} \\
\braket{p_+(t)}&=\sqrt{2 \hbar m \omega}(a_I \cos \Omega_+ t -K_+ a_R \sin \Omega_+ t)\\
\Delta x_+^2(t)&= \frac{\hbar}{2m \omega} = \Delta x_-^2(t)\\
\braket{x_-(t)} &= \sqrt{\frac{2 \hbar}{m \omega}} (b_R \cos \Omega_- t +  \frac{1}{K_-} b_I \sin \Omega_- t) \label{eq:simi22} \\ 
\braket{p_-(t)}&=\sqrt{2 \hbar m \omega}(b_I \cos \Omega_- t -K_- b_R \sin \Omega_- t)
\end{align}
\end{subequations}

 Applying the same reasoning that reduced equations \eqref{eq:qg_eom} to \eqref{eq:eom_rwa} here will similarly alter \eqref{eq:sc_eom}, thereby leading to the states being swapped at $t = \pi/(2 \omega_g)$. We will return to this point, but we shall first comment on the physics of equations \eqref{eq:sc_eom} and discuss some parallels to the RWA.
 
 The immediate point to note is that the first moments are identical for $H^q$ and $H^{sc}$, a fact well recognized in studies of semi-classical gravity\cite{yangMacroscopicQuantumMechanics2013}. The second moments do differ, but for $H^q$ they turn out to be constant for coherent states under the assumption $\omega_g \ll \omega$. For $H^{sc}$, they are already constant without this assumption, allowing the wave function to be calculated in closed-form, directly from \eqref{eq:sc_eom}.
 
 A second remark, for didactic purposes, pertains to situations where the oscillation amplitudes are large, or the experimental apparatus is sensitive enough to the $\delta$ factor corrections to the amplitudes. In such cases, the approximation $K_\pm \approx 1$ that we've made would not be appropriate. The state evolution for both the QG and SCEG cases would instead read:
\begin{align}\label{eq:sc_evo}
    \ket{\alpha}_1 \otimes \ket{\beta}_2 \mapsto  &\Ket{\rme^{-\rmi\omega t}(\alpha \cos \omega_g  t -\rmi \beta \sin \omega_g  t) - \delta A_t}_1 \otimes \nnl
    &\Ket{\rme^{-\rmi\omega t}(\beta \cos \omega_g  t -\rmi \alpha \sin \omega_g  t)  - \delta B_t}_2  
\end{align}
where $A_t =\rmi (a^* \sin \omega_+ t - b^* \sin\omega_- t)/2$ and $B_t = \rmi (a^* \sin \omega_+ t + b^* \sin\omega_- t)/2$. As expected, \eqref{eq:rwa_evo} is reproduced by ignoring the $\delta$ corrections to the amplitude of oscillations. However, for values of $|A_t|, |B_t|$ far exceeding $|1/\delta|$, the correction would be significant enough to warrant inclusion. The evolution predicted by \eqref{eq:sc_eom} and \eqref{eq:eom_rwa} then, would differ from \eqref{eq:rwa_evo}, which was obtained using the RWA Hamiltonian. This is in fact, an inadequacy of the RWA formalism, echoing an issue first identified by Walls~\cite{wallsHigherOrderEffects1972} in the context of quantum optics. It was shown there and elsewhere~\cite{wangPhotonDressedBlochSiegertShift2020} that measurements of the Bloch-Siegert shift increasingly undermine the RWA (resulting from the condition $\omega_g \gg \omega$ alone) as the photon number rises. A perturbative approach~\cite{puriMathematicalMethodsQuantum2001} later argued that the condition $\omega_g \braket{N} \gg \omega$ provides better results in this regime. 

This insight emerges naturally in our approach from equation \eqref{eq:sc_evo} when one notes that $\sqrt{\braket{N}} = |\alpha|$ and $A_t$ is a stand-in for $\alpha$. It follows that the RWA prediction \eqref{eq:rwa_evo} would agree with \eqref{eq:sc_evo}---the more accurate prediction---only if $\omega_g |A_t| \ll \omega$ holds.  Conversely, if $ \omega_g |A_t| \gtrsim \omega $, then the correction $\delta A_t$ would be sizeable, and \eqref{eq:sc_evo} \emph{would} differ from \eqref{eq:rwa_evo}, with the latter being invalid. Regarding the swapping of states via gravity, this is not a relevant concern in realistic experimental situations because $\delta \propto 1/d^3$ by definition and $d \gg |x_i -\braket{x_j}|$ (or $ d \gg |x_1 -x_2|$ for the QG case) by assumption. As such, we work in the regime $|A_t| \ll d $, ensuring that the correction in \eqref{eq:sc_evo} cannot be arbitrarily blown up. This in turn ensures that both the semi-classical evolution and the quantum gravity one always produce a swapping of initial coherent states, even for larger $A_t$, $B_t$.

\section{Discussion}

Starting from the Hamiltonian for the two-particle interaction we have shown that semiclassical gravity, just like quantized gravity, predicts state swapping of coherent states---a feature assumed to serve as an indicator for whether or not the gravitational interaction can be modeled by a LOCC channel.

An important remark is that, precisely speaking, not only is SCEG not a LOCC theory---it is \emph{not a channel} in the sense defined in the introduction. More specifically, SCEG defines a map $\Lambda : \mathrm{Proj}(\mathcal{H}) \to \mathrm{Proj}(\mathcal{H})$ between \emph{pure states}, i.e., within the space $\mathrm{Proj}(\mathcal{H})$ of projectors on one-dimensional subspaces of $\mathcal{H}$. For classical mixtures of Hilbert space states, it is a well-established fact~\cite{gisinStochasticQuantumDynamics1989,bassiNofasterthanlightsignalingImpliesLinear2015,giuliniCouplingQuantumMatter2022} that nonlinear dynamics do not define a map between density operators, as they can map equivalent ensembles (with the same density operator) to inequivalent ones. What we have shown is that among those allowed maps between pure states there are also such that, approximately, generate a swap as defined in equation~\eqref{eqn:swap}.

As far as experimental tests are concerned, given the numbers cited in subsection~\ref{sec:qg-full}, trapped-ion oscillators require unreasonably large times ($1/\omega_g > \SI{e10}{\second}$ or $\sim100$~years) to exhibit the effect we seek. Torsion pendulums offer the possibility of a larger $\omega_g$ and a shorter run time for the experiment, but the variations in the mean positions of the masses are estimated by Lami et al.~\cite{lamiTestingQuantumnessGravity2024} to be on the order of \SI{0.1}{\pico\meter}, which is already challenging to detect. The $\delta$ factor corrections to the amplitude of the oscillations may therefore be neglected still.

Strictly speaking, the equations of semi-classical gravity do not produce a swap unless we ignore the $\mathcal{O}(|\alpha|\delta)$ correction, but the same concession is needed for the QG case. No  experiment designed to observe the swap predicted by QG would detect any deviations from the swap map \eqref{eq:swap} should gravity be classical, because discerning the QG evolution in the RWA \eqref{eq:rwa_evo} from the semiclassical equations~\eqref{eq:sc_evo} mandates a precise measurement of the first moments. These are identical for QG and SCEG, as seen from \eqref{eq:qg_eom} and \eqref{eq:sc_eom}. Regarding Lami et al.'s work, the SCEG bypasses their LOCC theorems due to its nonlinearity, which satisfies a caveat discussed in their paper: ``gravity somehow `knows' the initial state of the system $\psi_\alpha$ and uses this knowledge to bypass the
above theoretical bounds, reproducing at the output
the correct state with high fidelity''~\cite{lamiTestingQuantumnessGravity2024}. Therefore, the swapping of a pair of coherent states is insufficient evidence on its own to establish that gravity is a quantum force (in the sense of perturbative QG) rather than classical (in the sense of SCEG). If one would maintain the definition that equates ``quantumness'' with a non-LOCC channel, then one must assert that SCEG is a quantum theory in that sense. Arguably, this definition contrasts most people's intuition that would consider the spacetime structure described by the semiclassical Einstein equations as evidently classical.

There is, however, a possible test in the spirit of these ideas that may be able to distinguish SCEQ from QG, which involves superpositions of coherent states. Consider the initial state
\begin{align*}
      \ket{\psi_0}=   \frac{\ket{\alpha}_1 + \ket{-\alpha}_1}{\sqrt{2}}\otimes \ket{0}_2 
\end{align*}
which evolves under $H^q_{\text{RWA}}$ to the following state:

\begin{align*}
    \ket{\psi_t} &= \ket{\alpha_t \cos \omega_g t }_1\ket{- \rmi \alpha_t \sin \omega_g t}_2 
+ \\&\ket{-\alpha_t \cos \omega_g t}_1 \ket{\rmi \alpha_t \sin \omega_g t}_2
\end{align*}
This result is deduced from linearity, and thus does not follow for SCEG due to its intrinsic non-linearity. Nevertheless, a key insight may be drawn by inspecting the EoM: that the first moments are all identically zero. While insignificant for the QG case, this means for the SCEG Hamiltonian that all interaction terms $x_i\braket{x_i}$ vanish, precluding any possibility of information transfer except for a shift in the oscillator frequency (which is inconsequential). This reaffirms the preconceived notion that SCEG cannot generate entanglement, at least when expanded to the 2nd order as in \eqref{eq:H_sc}. However, it raises the question of why this is the case despite the manifest non-locality of semiclassical gravity, and its ability to effectively mimic the non-LOCC the swap map--- as we've shown in this paper. An intuitive answer lies in the classicality of coherent states and the non-classical nature of their superpositions. Nevertheless, it would be an interesting challenge to mathematically demonstrate that the SCEG Hamiltonian indeed always maps a product state to another product state. 

To conclude, we would like to emphasize that, despite its unfavorable properties (in particular the apparent possibility for superluminal signaling), the Schr{\"o}dinger-Newton equation has some merit and a definite regime of validity.  While there exist some hybrid models that better approximate QG, even to the extent of generating entanglement, they all entail ad-hoc supplements such as a free parameter or a post-quantum ontology. The complete lack of such frails adds to the enduring intrigue of SCEG as a phenomenological theory---if not a fundamental one--- that continues to elude experimental refutation despite its theoretical challenges.

\begin{acknowledgments}
All authors acknowledge funding through the Volkswagen Foundation.
\end{acknowledgments}

\appendix

\section{Equations of Motion}\label{App:EoM}
To arrive at the equations of motion presented in the main text, we study the following template Hamiltonian:
\begin{equation}
    H_0 = A p^2 + B x^2 + C\ev{x}x \,.
\end{equation}
All Hamiltonians discussed in this paper fit this general form in the normal mode coordinates, meriting a derivation of their EoM. 
In the Heisenberg picture, the evolution of operators $O$ with no explicit time dependence is given by 
\begin{equation}
\frac{\rmd O}{\rmd t} = \frac{\rmi}{\hbar} [H_0,O] \,.    
\end{equation}

\begin{itemize}
    \item  For $x$, we have
    \begin{align*}
    \frac{\rmd x}{\rmd t} &= \frac{\rmi}{\hbar}(A[p^2,x] + B[x^2,x]+C \braket{x}[x,x])\\
    &=\frac{\rmi}{\hbar}[A(-2 \hbar i p) + 0 + 0]\\
    &= 2 A p
    \end{align*}

    \item For $p$,
    \begin{align*}
    \frac{\rmd p}{\rmd t} &= \frac{\rmi}{\hbar}(A[p^2,p] + B[x^2,p]+C \braket{x}[x,p])\\
    &=\frac{\rmi}{\hbar}[0 + B\left(-\frac{2\hbar x}{i}\right) + C \braket{x}i \hbar]\\
    &= -2 B x - C \braket{x}
    \end{align*}

    \item For $x^2$,
    \begin{align*}
    \frac{\rmd x^2}{\rmd t} &= \frac{\rmi}{\hbar}(A[p^2,x^2] + B[x^2,x^2]+C \braket{x}[x,x^2])\\
    &=\frac{\rmi}{\hbar}[A(-4 i \hbar x p - 2 \hbar^2) + 0 + 0]\\
    &= 2 A(xp+px)
    \end{align*}

    \item For $p^2$,
    \begin{align*}
    \frac{\rmd p^2}{\rmd t} &= \frac{\rmi}{\hbar}(A[p^2,p^2] + B[x^2,p^2]+C \braket{x}[x,p^2])\\
    &=\frac{\rmi}{\hbar}[0 + B(4 i \hbar x p + 2 \hbar^2) + C\braket{x}(2 \hbar i p) ]\\
    &= -2 B(xp+px)- 2 C \braket{x} p
    \end{align*}

    \item Next, we consider the evolution of $xp+px$ and note that
    \begin{align*}
    xp + px &= x p + xp - i \hbar\\
    \Rightarrow \frac{\rmd (xp + px)}{\rmd t} &= 2 \frac{\rmd \: xp}{\rmd t}
    \end{align*}
    
    The derivative on the right works out to
    \begin{align*}
    \frac{\rmd \: xp}{\rmd t} &= \frac{\rmi}{\hbar}(A[p^2,xp] + B[x^2,xp]+C \braket{x}[x,xp])\\
    &=\frac{\rmi}{\hbar}[A(-2 \hbar i p^2) + B\left(\frac{-2 \hbar x^2}{i}\right) + C\braket{x}(i \hbar x) ]\\
    &= 2 A p^2-2 Bx^2 - C \braket{x} x\\
    \end{align*}
    \end{itemize}
    With these operator derivatives, by virtue of the Ehrenfest theorem, we obtain the time evolution for the variances:
    \begin{itemize}
    \item For $V_{xx} = \braket{x^2}-\braket{x}^2$, 
    \begin{align*}
        \frac{\rmd V_{xx}}{\rmd t}&= \frac{\rmd \: \braket{x^2}}{\rmd t} - \frac{\rmd }{\rmd t}(\braket{x}^2)\\
        & = \frac{\rmd \: \braket{x^2}}{\rmd t} - 2 \braket{x}\frac{\rmd  \braket{x}}{\rmd t}\\
        & = 4 A \left(\frac{\braket{xp +px}}{2} - \braket{x} \braket{p}\right)\\
        \frac{\rmd  V_{xx}}{\rmd t}&= 4 A V_{xp}
    \end{align*}

    \item For $V_{pp} = \braket{p^2}-\braket{p}^2$, 
    \begin{align*}
        \frac{\rmd  V_{pp}}{\rmd t}&= \frac{\rmd \: \braket{p^2}}{\rmd t} - \frac{\rmd  }{\rmd t}(\braket{p}^2)\\
        & = \frac{\rmd \: \braket{p^2}}{\rmd t} - 2 \braket{p}\frac{\rmd  \braket{p}}{\rmd t}\\
        & = -4 B \left(\frac{\braket{xp +px}}{2} - \braket{x} \braket{p}\right)\\
        \frac{\rmd  V_{pp}}{\rmd t}&= -4B V_{xp}
    \end{align*}

    \item For $V_{xp} = \dfrac{\braket{xp+px}}{2} - \braket{x}\braket{p}$,
    \begin{align*}
        \frac{\rmd  V_{xp}}{\rmd t}&= \frac{\rmd }{\rmd t} \frac{\braket{xp + px}}{2} - \frac{\rmd  }{\rmd t}(\braket{x}\braket{p})\\
        & = \frac{1}{2}\frac{\rmd  \braket{xp+px}}{\rmd t} - \braket{p}\frac{\rmd  \braket{x}}{\rmd t} -\braket{x}\frac{\rmd  \braket{p}}{\rmd t}\\
        & = 2 A(\braket{p^2}-\braket{p}^2) -2B(\braket{x^2}-\braket{x}^2)\\
        \frac{\rmd  V_{xp}}{\rmd t}&= 2 A V_{pp}-2B V_{xx}
    \end{align*}

\end{itemize}
Note that the evolution of these second moments is completely independent of the linear coefficient $C$.

We can now apply these results to the three Hamiltonians of interest, where we assume initial conditions corresponding to a product of two coherent states; as shown in equations~\eqref{eq:cs_tran}, a product state in the original coordinates remains a product state in the normal mode coordinates $x_+,x_-$. The initial conditions, common to all three cases to be discussed, are:
\begin{align}\label{eqn:app-initial-cond}
 \braket{x_+(0)} &= \frac{1}{\sqrt{2}} \left( \sqrt{\frac{2 \hbar}{m \omega}} \alpha_R   + \sqrt{\frac{2 \hbar}{m \omega}} \beta_R \right)\\
 & = \sqrt{\frac{2 \hbar}{m \omega}}a_R  \\
\braket{p_+(0)}&= \sqrt{2 \hbar m \omega} \left(\frac{\alpha_I}{\sqrt{2}}  + \frac{\beta_I }{\sqrt{2}}\right)\\
& = \sqrt{2 \hbar m \omega} b_I\\
\Delta x_+^2(0)&= \frac{1}{2}(\Delta x_1^2 +\Delta x_2^2) =\Delta x_-^2(0)\\
&= \frac{\hbar}{2m \omega}\\
\Delta p_+^2(0) &= \frac{1}{2}(\Delta p_1^2 +\Delta p_2^2)\\
& = \frac{m \hbar \omega}{2} = \Delta p_-^2(0)\\
\braket{x_-(0)} &= \sqrt{\frac{2 \hbar}{m \omega}}\left(    \frac{\alpha_R}{\sqrt{2}}  - \frac{\beta_R}{\sqrt{2}}  \right)\\
 & = \sqrt{\frac{2 \hbar}{m \omega}}b_R  \\
\braket{p_-(0)}&= \sqrt{2 \hbar m \omega} \left(\frac{\alpha_I}{\sqrt{2}} - \frac{\beta_I }{\sqrt{2}}\right)\\
\end{align}

\paragraph{Quantum Gravity with Rotating Wave Approximation}
with the approximate effective Hamiltonian
\begin{equation*}
    H^q_{\text{RWA}} = \sum_{j=+,-}k_j\left(\frac{p_j^2}{2m} + \frac{1}{2} m \omega^2 x_j^2\right)  
\end{equation*}
produces the following  evolution of first and second moments:
\begin{align*}
    \frac{\rmd  \braket{x_\pm}}{\rmd t} & = \frac{k_\pm\braket{p_\pm}  }{m}\\
    \frac{\rmd \braket{p_\pm}}{\rmd t}&= -m k_\pm \omega^2 \braket{x_\pm}\\
\dot{V}_{x_\pm x_\pm} &= \frac{2k_\pm}{m}V_{x_\pm p_\pm} \\ 
\dot{V}_{p_\pm p_\pm}&= -2k_\pm m \omega^2 V_{x_\pm p_\pm}\\
\dot{V}_{x_\pm p_\pm} & = \frac{k_\pm}{m}V_{p_\pm p_\pm} - m k_\pm\omega^2 V_{x_\pm x_\pm}
\end{align*}
Solving these EoM, we find
\begin{align*}
\braket{x_\pm(t)} &= \braket{x_\pm(0)} \cos \omega_\pm t + \frac{\braket{p_\pm(0)}}{ m\omega} \sin \omega_\pm t \\
\braket{p_\pm(t)}&=\braket{p_\pm(0)} \cos \omega_\pm t - m\omega \braket{x_\pm(0)} \sin \omega_\pm t\\
\Delta x_\pm^2(t)& = \Delta x_\pm^2(0)\cos^2\omega_\pm t + \frac{\Delta p_\pm^2(0)}{m^2 \omega^2} \sin^2 \omega_\pm t\\ 
& \quad + \frac{\sin 2\omega_\pm t}{m \omega} V_{x_\pm p_\pm}(0) 
\end{align*}
Substituting the initial conditions \eqref{eqn:app-initial-cond}, we obtain equations~\eqref{eq:eom_rwa}.

\paragraph{Full Quantum Gravity}
with the effective Hamiltonian
\begin{equation*}
    H^q=\sum_{j=+,-} \left[\frac{p_j^2}{2 m} + \frac{1}{2} m\Omega_j^2 x_j^2 \right]
\end{equation*}
results in the following equations:
\begin{align*}
\frac{\rmd  \braket{x_\pm}}{\rmd t} & = \frac{\braket{p_\pm}  }{m}\\
\frac{\rmd \braket{p_\pm}}{\rmd t}&= -m \Omega_\pm^2 \braket{x_\pm}\\
\dot{V}_{x_\pm x_\pm} &= \frac{2}{m}V_{x_pm p_\pm} \\ 
\dot{V}_{p_\pm p_\pm} &= -2m \Omega_\pm^2 V_{x_\pm p_\pm}\\
\dot{V}_{x_\pm p_\pm} & = \frac{V_{p_\pm p_\pm}}{m} - m \Omega_\pm^2 V_{x_\pm x_\pm}
\end{align*}
Solving these EoM, we have
\begin{align*}
\braket{x_\pm(t)} &= \braket{x_\pm(0)} \cos \Omega_\pm t + \frac{\braket{p_\pm(0)}}{ m\Omega_\pm} \sin \Omega_\pm t\\
\braket{p_\pm(t)}&=\braket{p_\pm(0)} \cos \Omega_\pm t - m\Omega_\pm \braket{x_+(0)} \sin \Omega_\pm t\\
\Delta x_\pm^2(t)& = \Delta x_\pm^2(0)\cos^2\Omega_\pm t + \frac{\Delta p_\pm^2(0)}{m^2 \Omega_\pm^2} \sin^2 \Omega_\pm t\\ & \quad + \frac{\sin 2\Omega_\pm t}{m \Omega_\pm} V_{x_\pm p_\pm}(0) 
\end{align*}

\paragraph{Semiclassical Gravity}
with the Hamiltonian
\begin{equation*}
    H^{sc} = \sum_{j=+,-} \left[\frac{p_j^2}{2 m} + \frac{1}{2} m\omega^2 x_j^2 \right]+ 2C x_+ \braket{x_+} - 2C x_- \braket{x_-}
\end{equation*}
results in
\begin{align*}
\frac{\rmd  \braket{x_\pm}}{\rmd t} & = \frac{\braket{p_\pm}  }{m}\\
\frac{\rmd \braket{p_\pm}}{\rmd t}&= -m \Omega_\pm^2 \braket{x_\pm}\\
\dot{V}_{x_\pm x_\pm} &= \frac{2}{m}V_{x_\pm p_\pm} \\ 
\dot{V}_{p_\pm p_\pm} &= -2m \omega^2 V_{x_\pm p_\pm}\\
\dot{V}_{x_\pm p_\pm}& = \frac{V_{p_\pm p_\pm}}{m} - m \omega^2 V_{x_\pm x_\pm}
\end{align*}
Solving these EoM yields
\begin{align*}
\braket{x_\pm(t)} &= \braket{x_\pm(0)} \cos \Omega_\pm t + \frac{\braket{p_\pm(0)}}{ m\Omega_\pm} \sin \Omega_\pm t\\
\braket{p_\pm(t)}&=\braket{p_\pm(0)} \cos \Omega_\pm t - m\Omega_\pm \braket{x_+(0)} \sin \Omega_\pm t\\
\quad \quad\Delta x_\pm^2(t)& = \Delta x_\pm^2(0)\cos^2\omega t + \frac{\Delta p_\pm^2(0)}{m^2 \omega^2} \sin^2 \omega t\\
& \quad + \frac{\sin 2\omega t}{m \omega} V_{x_\pm p_\pm}(0) 
\end{align*}
Substituting initial values leads to \eqref{eq:sc_eom}.

\section{Wave Function Evolution} \label{App:WFE}
The time evolution of the initial product of coherent states follows directly from the evolution of the operators as derived in appendix~\ref{App:EoM}. Consider a coherent state $\ket{\alpha}$, where $\alpha = \alpha_R + \rmi \alpha_I$ is a complex number. The expectation values of the position and momentum are, respectively,

\begin{align}
    \braket{\alpha | \hat{x} | \alpha}&= \sqrt{\frac{\hbar}{ 2 m \omega}}(\alpha + \alpha^*) = \sqrt{\frac{2 \hbar}{m \omega}} \alpha_R\\
    \braket{\alpha | \hat{p} | \alpha}&= -\rmi\sqrt{\frac{\hbar m \omega}{ 2}}(\alpha - \alpha*) = \sqrt{2 \hbar m \omega} \alpha_I
\end{align}
\quad We see that we can deduce the parameter $\alpha$ entirely from these two measurements. We can leverage this simple observation to obtain the time evolution of a coherent state under any Hamiltonian that preserves its coherence (width). To this end, we calculate the evolution of $\hat{x}(t)$ and $\hat{p}(t)$ in the Heisenberg picture, inspect $\braket{\hat{x}(t)}$ and $\braket{\hat{p}(t)}$ and infer $\alpha_t$ thence. 

We now demonstrate this technique using the Hamiltonian in \eqref{eq:H_sc}
$$
H^{sc}=\sum_{j=+,-} \left[\frac{p_j^2}{2 m} + \frac{1}{2} m\omega^2 x_j^2 \right]+ 2\lambda x_+ \braket{x_+} - 2\lambda x_- \braket{x_-}
$$
since it is the most general one in this paper.

From equations~\eqref{eq:eom_rwa} we have
\begin{align}
\braket{x_+(t)} &= \sqrt{\frac{2 \hbar}{m \omega}} (a_R \cos \Omega_+ t +  \frac{1}{K_+} a_I \sin \Omega_+ t) \label{eqn:app-evx}\\
\braket{p_+(t)}&=\sqrt{2 \hbar m \omega}(a_I \cos \Omega_+ t -K_+ a_R \sin \Omega_+ t) \label{eqn:app-evp}\\
\Delta x_+^2(t)&= \frac{\hbar}{2m \omega} \label{eqn:app-delta-x-squared}
\end{align}

From \eqref{eqn:app-delta-x-squared}, we see that the width remains constant in time, implying that coherent states remain coherent under this Hamiltonian. For the initial state $\ket{a}_+ \otimes \ket{b}_-$, the terms within the paranthesis of \eqref{eqn:app-evx} and \eqref{eqn:app-evp} determine, respectively, the real and imaginary parts of $a(t)$. Constructing the complex number $a(t)$, we have
\begin{align*}
 a(t)=  (a_R +i a_I) \cos \Omega_+ t +  \left(\frac{a_I }{K_+} - i K_+ a_R \right) \sin \Omega_+ t
\end{align*}
We now approximate $K_\pm \approx 1 \pm \delta$ and obtain
$$
a(t) = a \rme^{-\rmi\omega_+ t} -  i\delta a^*\sin \omega_+ t 
$$

A similar expression follows for $b$:
$$
b \rightarrow b(t) = b \rme^{-\rmi \omega_- t} + \rmi\delta b^* \sin\omega_- t
$$
The time evolution therefore reads
\begin{align*}
    \ket{a}_+ \otimes \ket{b}_- \rightarrow& \Ket{a \rme^{-\rmi\omega_+ t} -  \rmi\delta a^*\sin \omega_+ t }_+ \\
    &\otimes \Ket{b \rme^{-\rmi \omega_- t} + \rmi\delta b^* \sin\omega_- t}_- \,.
\end{align*}

\end{document}